\documentclass[preprint,epsfig,onecolumn,floats,showpacs]{revtex4}
\usepackage{amssymb}
\usepackage{graphicx}
\usepackage{epsfig}

\begin{document}

\title{Always On Non-Nearest-Neighbor Coupling in Scalable Quantum Computing}
\author{Yong Hu}
\email{yhu3@mail.ustc.edu.cn}
\affiliation{Key Laboratory of Quantum Information, University of Science and Technology
of China, Chinese Academy of Sciences, Hefei, Anhui 230026, China}
\author{Zheng-Wei Zhou}
\email{zwzhou@ustc.edu.cn}
\affiliation{Key Laboratory of Quantum Information, University of Science and Technology
of China, Chinese Academy of Sciences, Hefei, Anhui 230026, China}
\author{Guang-Can Guo}
\email{gcguo@ustc.edu.cn}
\affiliation{Key Laboratory of Quantum Information, University of Science and Technology
of China, Chinese Academy of Sciences, Hefei, Anhui 230026, China}

\begin{abstract}
We study the non-nearest-neighbor interaction effect in $1$-D spin-$1/2$
chain model. In many previous schemes this long-range coupling is omitted
because of its relative weak strength compared with the nearest-neighbor
coupling. We show that the quantum gate deviation induced by the omitted
long-range interaction depends on not only its strength but also the scale
of the system. This implies that omitting the long-range interaction may
challenge the scalability of previous schemes. We further propose a quantum
computation scheme. In this scheme, by using appropriate encoding method, we
effectively negate influence of the next-nearest-neighbor interaction in
order to improve the precision of quantum gates. We also discuss the
feasibility of this scheme in $1$-D Josephson charge qubit array system.
This work may offer improvement in scalable quantum computing.
\end{abstract}

\maketitle

\section{Introduction}

One of the critical problems in realizing scalable quantum computation (QC)
is performing two qubits gates, which implies that the couplings between
qubits are variable functions subject to external control. In many physical
systems, this requirement is not easy to be satisfied. Recently various
\textquotedblleft always on\textquotedblright\ QC schemes have been proposed
to solve this problem. Zhou \textit{et al.} suggest encoding logical qubits
in interaction free subspace (IFS) \cite{Zhou1}\cite{Zhou2}, while Benjamin
\textit{et al. }suggest changing interaction type between qubits from
non-diagonal Heisenberg type to diagonal Ising type by tuning the Zeeman
energy splits of individual qubits \cite{Bose1}. Schemes implementing above
ideas into AFM spin ring and optical systems have already been developed
\cite{Troiani}\cite{Lovett}.

These above elegant proposals mainly concern a $1$-D spin-$1/2$ chain with
couplings between neighboring qubits. This model is a good analogue to many
candidate scalable QC implementations. But in realistic systems such as
quantum dot and optical lattice, there is not only the nearest-neighbor
interaction, but also the next-nearest-neighbor, or even longer range
interactions existing. In this paper we study effect of the
non-nearest-neighbor coupling in $1$-D spin-$1/2$\ chain model. Since the
strength of the long-range coupling is often much smaller than that of the
nearest-neighbor coupling, in many previous schemes its effect was ignored.
But this omission results deviation in performing quantum gates. In this
paper first we estimate the deviation of quantum gates induced by the
omitted long-range interaction. Results imply that this deviation depends on
not only the long-range interaction strength but also the length of the spin
chain. This means that though the long-range interaction strength is very
small, the induced deviation can \textquotedblleft
cumulate\textquotedblright\ on the whole $1$-D spin-$1/2$ chain, hence
scalability of previous QC scheme is restricted. Following this estimation,
we consider how to suppress this unwanted effect. We propose a QC scheme in
which by using proper encoding methods, influence of the perpetual
next-nearest-neighbor interaction is effectively negated, hence the
precision of quantum gates updated. Compared with previous schemes in which
the long-range coupling was omitted, our scheme does not cause the speed of
quantum gates slow evidently. We also discuss the feasibility of this scheme
in $1$-D Josephson charge qubit array system.

This paper is organized as follow: In the second section we consider a $1$-D
spin-$1/2$ chain model. In the recent \textquotedblleft
always-on\textquotedblright\ QC schemes \cite{Zhou2}\cite{Bose1}, long-range
interaction terms are always neglected. Here, we estimate the quantum gates
deviation induced by an omitted perpetual next-nearest-neighbor Ising type
interaction term. In the third section we consider how to suppress influence
of the untunable long-range coupling. We use \textquotedblleft blockade
spin\textquotedblright\ encoding method to effectively neutralize unwanted
influence of the perpetual next-nearest-neighbor interaction. Based on this
encoding method we propose a QC scheme. In the fourth section we study the
feasibility of this scheme in $1$-D Josephson junction charge qubit array
system. Several potential generalizations are suggested before conclusion.

\section{influence of the permanent long-range interaction}

We start with a $1$-D spin-$1/2$\ chain consisting of $2n+1$ spins, with
tunable $XXZ$ interaction between neighboring spin \cite{Wu2}. The
Hamiltonian is:
\begin{equation}
H_{Ideal}=H_{S}+H_{I}  \label{HamiltonianIdea1}
\end{equation}%
where
\begin{equation}
H_{S}=\sum_{i=1}^{2n+1}H_{S}^{i}=\sum_{i=1}^{2n+1}B_{x}^{i}\sigma
_{i}^{x}+B_{z}^{i}\sigma _{i}^{z}  \label{HamiltonianIdea2}
\end{equation}%
\begin{equation}
H_{I}=\sum\limits_{i=1}^{2n}H_{I}^{i,i+1}=\sum\limits_{i=1}^{2n}J_{i,i+1}(%
\sigma _{i}^{x}\sigma _{i+1}^{x}+\sigma _{i}^{y}\sigma
_{i+1}^{y})+J_{1}\sigma _{i}^{z}\sigma _{i+1}^{z}  \label{HamiltonianIdea3}
\end{equation}%
We assume that only values of $B_{x}^{i}$, $B_{z}^{i}$, and $J_{i,i+1}$ are
tunable, while $J_{1}$ remains constant. Various systems including single
electron arrays, optical lattice, quantum dot, and Josephson junction array
could be described by this \textquotedblleft popular\textquotedblright\
Hamiltonian, while methods developed for the $XXZ$ type exchange interaction
can be easily generalized to other types of exchange interaction including $%
XY$ type and Heisenberg type interaction \cite{Bose1}.

But in realistic systems there is not only the nearest-neighbor coupling,
but also residual non-nearest-neighbor couplings existing. The interaction
terms between non-neighboring qubits may origin from residual wave functions
overlap or long-range Coulomb interaction. Without loss of generality, we
may first estimate the influence from the next-nearest-neighbor coupling. We
could assume that a permanent\ next-nearest-neighbor interaction term is
ignored in Eq. \ref{HamiltonianIdea1}:
\begin{equation}
H_{L}=J_{2}\sum_{i=1}^{2n-1}\sigma _{i}^{z}\sigma _{i+2}^{z}
\label{LongRange Interaction}
\end{equation}%
Thus though the theoretical evolution of the system is governed by $H_{Ideal}
$, the realistic evolution is governed by the realistic Hamiltonian $%
H_{R}=H_{Ideal}+H_{L}$. When we perform quantum gates following previous
schemes in which only the nearest-neighbor interaction is concerned \cite%
{Zhou2}\cite{Bose1}\cite{Wu2}, deviation of the realistic evolution from the
ideal expectation is induced by $H_{L}$.

Let us follow schemes developed in ref. \cite{Bose1} as an example. For
untunable $\sum\limits_{i=1}^{2n}J_{1}\sigma _{i}^{z}\sigma _{i+1}^{z}$ term
in $H_{I}$, methods of freezing \textquotedblleft
blockade\textquotedblright\ spins in definite states between logical qubits
have been proposed to negate the influence of this always-on coupling on
single qubit operations. As shown in Fig. 1: In the spin-$1/2$ chain, only
the even spins (the hexagonal ones) are chosen as qubits while the odd spins
(the rounded ones) are used as \textquotedblleft blockade\textquotedblright
. When performing single qubit operations, we would make the blockade spins
\textquotedblleft frozen\textquotedblright\ in states $\left\vert
0\right\rangle $ and $\left\vert 1\right\rangle $ as shown in Fig. 1, thus
the influence of the perpetual $\sum\limits_{i=1}^{2n}J_{1}\sigma
_{i}^{z}\sigma _{i+1}^{z}$ terms on the even spins is effectively
neutralized because $J_{1}\sigma _{2i}^{z}(\sigma _{2i-1}^{z}+\sigma
_{2i+1}^{z})=0$ for any $i$. So there are $n$ qubits and $n+1$
\textquotedblleft blockade spins\textquotedblright\ in the whole chain,
where the $i$th qubit is encoded by the $2i$th spin. The single qubit gates
on the $i$th qubit are realized by tuning the effective magnetic field on
the $2i$th spin, while two qubits gates between the $i$th qubit and the $%
\left( i+1\right) $th qubit could be established by tuning the inter-spin
exchange interaction strength $J_{2i,2i+1}$ and $J_{2i+1,2i+2}$.

\begin{figure}[tbh]
\epsfig{file=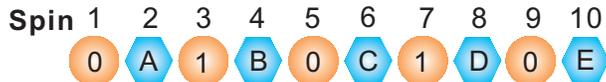,width=8cm}
\caption{For the always-on $\sum\limits_{i=1}^{2n}J_{1}\protect\sigma %
_{i}^{z}\protect\sigma _{i+1}^{z}$ type inter-spin coupling, qubits will
suffer continuous phase gates with their neighbors. We may only use the even
spins (the hexagonal ones) to encode information while use the odd spins
(the rounded ones) as blockade spin. When performing single qubit operation,
the blockade spins are \textquotedblleft frozen\textquotedblright\ in
definite states $\left\vert 0\right\rangle $ or $\left\vert 1\right\rangle $
in order to negate influence of the perpetual Ising interaction.}
\end{figure}

When we perform quantum gates, the theoretical expected unitary evolution of
the whole chain is $U=\exp (-itH_{Ideal})$ while the realistic evolution is $%
V=\exp [-it(H_{Ideal}+H_{L})]$. We could estimate the deviation
$\left\Vert U-V\right\Vert $ in single and multi qubit operations,
where for convenience we use the definition of spectral norm, that
is, the norm of an operator $O$ is defined as the square root of
the maximum eigenvalue of $O^{\dagger }O$. As shown in appendix.
A, when we perform a certain quantum gate, the deviation induced
by the omitted next-nearest-neighbor interaction is at least of
the order $\left\vert \exp \{-iJ_{2}tn\}-1\right\vert $, $t$ being
the required time for performing this gate. For $t$ very small, we
could estimate the deviation speed:
\begin{equation}
\frac{d}{dt}\left\Vert U-V\right\Vert \backsim O(nJ_{2})
\label{DeviationSpeed}
\end{equation}%
This result implies that, though $J_{2}$ is very small, deviation induced by
$H_{L}$ could become very large because this deviation depends on the length
of the chain. This deviation could increase with the chain become longer.
Thus omitting the long-range interaction effect may restrict the scalability
of previous schemes. Below we consider how to handle this problem by using
proper encoding methods.

\section{\protect\bigskip Using Encoding Schemes To suppress the influence
of long-range interaction}

Let us illustrate our idea intuitively. Following the previous
section we start with a $1$-D spin-$1/2$ chain with tunable
nearest-neighbor $XXZ$ interaction and next-nearest-neighbor Ising
interaction. The Hamiltonian reads:
\begin{equation}
H_{M}=H_{S}+H_{I}+H_{L}  \label{modelhamiltonian1}
\end{equation}%
Where $H_{S}$, $H_{I}$, and $H_{L}$ are described by Eqs. \ref%
{HamiltonianIdea2}-\ref{LongRange Interaction}. Here, we further set $%
B_{z}^{i}=0$ for any $i$ in the whole quantum information process, i. e.
there is no $\sigma ^{z}$ terms in $H_{S}$.

As sketched in Fig. 2, we encode one logical qubit by two physical spins,
using the spin states $\left\vert 01\right\rangle $ and $\left\vert
10\right\rangle $ as logical $\left\vert 0\right\rangle $ and $\left\vert
1\right\rangle $. Thus the tunable $XY$ interaction between two spins plays
the role of $\sigma ^{x}$ rotation in the 2-D logical Hilbert space \cite%
{Wu2}.

Similar to the \textquotedblleft blockade spin\textquotedblright\ methods
used in ref. \cite{Bose1}, our intuitive idea is freezing two
\textquotedblleft blockade spins\textquotedblright\ between each two logical
qubits to negate the influence of the nearest-neighbor and
next-nearest-neighbor always-on Ising type interaction. As shown in Fig. 2:
The hexagonal ones are spins used to encode information while the octagonal
ones are used as blockade. We use the spins 3 and 4 as one logical qubit
while spins 7 and 8 as another. When performing single qubit operations, we
would make the blockade spins (spins 1, 2, 5, 6, 9, and 10) all
\textquotedblleft frozen\textquotedblright\ in state $\left\vert
0\right\rangle $, thus the influence of permanent Ising type interaction on
single logical qubit can be effectively negated: Since spin $3$-$4$ are in
Hilbert subspace spanned by spin states $\left\vert 01\right\rangle $ and $%
\left\vert 10\right\rangle $, we simply calculate the influence of
Ising type interactions on the logical qubit and get:
\begin{equation}
J_{1}(\sigma _{2}^{z}\sigma _{3}^{z}+\sigma _{4}^{z}\sigma
_{5}^{z})+J_{2}(\sigma _{1}^{z}\sigma _{3}^{z}+\sigma _{2}^{z}\sigma
_{4}^{z}+\sigma _{3}^{z}\sigma _{5}^{z}+\sigma _{4}^{z}\sigma _{6}^{z})=0
\end{equation}

\begin{figure}[tbh]
\epsfig{file=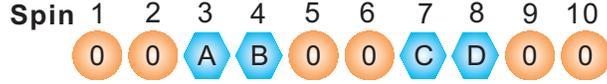,width=8cm}
\caption{We use two spins to encode one logic qubit, that is, the two spins
3-4 as one qubit and spins 7-8 as another. Two \textquotedblleft blockade
spins\textquotedblright\ both in $\left\vert 0\right\rangle $ state are
placed between two logical qubits in order to negate the influence of
next-nearest-neighbor interaction.}
\end{figure}

Now we show how to perform universal quantum gates, that is, the
single qubit $\sigma ^{x}$ rotations, the single qubit $\sigma
^{z}$ rotations, and CPHASE gates between two qubits. As shown in
Fig. 2, performing the single qubits $\sigma ^{x}$ rotations on
qubit encoded by spin 3 and 4 is easy by tuning $J_{3,4}$. We
further note a trivial fact mentioned by ref. \cite{Zhou3} that
the single qubit $\sigma ^{z}$ rotation can be constructed by the
CPHASE gate and single qubit $\sigma ^{x}$ rotation. So the
central problem becomes performing CPHASE gate between two logical
qubits. Our main idea to achieve this goal is, by tuning the
exchange interaction $J_{4,5}$ and $J_{6,7}$, we could perform a
CPHASE gate between logical qubits encoded by spin 3-4 and spin
7-8, adding a phase on one of the four logical qubits states while
remain the other three states unchanged.

We separate the spins 3, 4 and 5 as one party and spins 6, 7 and 8 as the
other. We set $B_{x}^{i}=0$ for any i in the whole process of performing two
qubit gate. Starting with the four possible initial states $\left\{
\left\vert 100\right\rangle ,\left\vert 010\right\rangle \right\}
_{3,4,5}\bigotimes \left\{ \left\vert 010\right\rangle ,\left\vert
001\right\rangle \right\} _{6,7,8}$, initially we tune all the strength of
nearest-neighbor $XY$ interaction to zero: $J_{i,i+1}=0$ for any i. Thus the
the four possible logical qubits states have equal static energy and we
define this static energy as energy zero point. As mentioned above, in
performing two qubit gate, what we may tune is just $J_{4,5}$ and $J_{6,7}$,
so we can reduce the Hamiltonian $H_{M}$ in Eq. \ref{modelhamiltonian1} into
a $9$-D Hilbert space%
\begin{equation}
S=span\{\left\{ \left\vert 100\right\rangle ,\left\vert 010\right\rangle
,\left\vert 001\right\rangle \right\} _{3,4,5}\bigotimes \left\{ \left\vert
001\right\rangle ,\left\vert 010\right\rangle ,\left\vert 100\right\rangle
\right\} _{6,7,8}\}  \label{WholeSpace}
\end{equation}%
The whole Hamiltonian $H_{M}$ becomes a function of tunable $J_{4,5}$ and $%
J_{6,7}$: $H_{M}=H_{M}(J_{4,5},J_{6,7})$. Below we use the label $\left\vert
abcdef\right\rangle $ to label the quantum states of the six spins from spin
3 to spin 8, the first $a$ for spin 3, the second $b$ for spin 4...... the
last $f$ for spin 8. For example, $\left\vert 100010\right\rangle $ labels
the state in which the spin 3 and spin 7 are in state $\left\vert
1\right\rangle $ while the spin 4, 5, 6 and 8 are in state $\left\vert
0\right\rangle $.

Indeed, $S$ in Eq. \ref{WholeSpace} can be reduced to four Hilbert
subspaces: The first one is a 1-D trivial subspace $S_{1}$ spanned by single
state $\left\vert 100001\right\rangle $. The second is 2-D
\begin{equation}
S_{2}=span\{\left\{ \left\vert 100\right\rangle \right\} _{3,4,5}\bigotimes
\left\{ \left\vert 010\right\rangle ,\left\vert 100\right\rangle \right\}
_{6,7,8}\}
\end{equation}%
Under basis $\left\{ \left\vert 100010\right\rangle ,\left\vert
100100\right\rangle \right\} $ Hamiltonian $H_{M}(J_{4,5},J_{6,7})$ can be
reduced to:%
\begin{equation}
H_{2}(J_{4,5},J_{6,7})=\left[
\begin{array}{cc}
0 & 2J_{6,7} \\
2J_{6,7} & 0%
\end{array}%
\right]
\end{equation}

The third is similar to the second:
\begin{equation}
S_{3}=span\{\left\{ \left\vert 010\right\rangle ,\left\vert 001\right\rangle
\right\} _{3,4,5}\bigotimes \left\{ \left\vert 001\right\rangle \right\}
_{6,7,8}\}
\end{equation}%
Under basis $\left\{ \left\vert 010001\right\rangle ,\left\vert
001001\right\rangle \right\} $ the reduced Hamiltonian is
\begin{equation}
H_{3}(J_{4,5},J_{6,7})=\left[
\begin{array}{cc}
0 & 2J_{4,5} \\
2J_{4,5} & 0%
\end{array}%
\right]
\end{equation}

The fourth is 4-D
\begin{equation}
S_{4}=\left\{ \left\vert 010\right\rangle ,\left\vert 001\right\rangle
\right\} _{3,4,5}\bigotimes \left\{ \left\vert 010\right\rangle ,\left\vert
100\right\rangle \right\} _{6,7,8}
\end{equation}%
Under basis $\left\{ \left\vert 010010\right\rangle ,\left\vert
010100\right\rangle ,\left\vert 001010\right\rangle ,\left\vert
001100\right\rangle \right\} $ the reduced Hamiltonian is%
\begin{equation}
H_{4}(J_{4,5},J_{6,7})=\left[
\begin{array}{cccc}
0 & 2J_{6,7} & 2J_{4,5} & 0 \\
2J_{6,7} & -4J_{2} & 0 & 2J_{4,5} \\
2J_{4,5} & 0 & -4J_{2} & 2J_{6,7} \\
0 & 2J_{4,5} & 2J_{6,7} & -4J_{1}%
\end{array}%
\right]
\end{equation}%
The form of $H_{4}(J_{4,5},J_{6,7})$ is quite similar to the NMR type
Hamiltonian: If we set $span\left\{ \left\vert 010\right\rangle ,\left\vert
001\right\rangle \right\} _{3,4,5}$ as the left \textquotedblleft
qubit\textquotedblright\ and $span\left\{ \left\vert 010\right\rangle
,\left\vert 100\right\rangle \right\} _{6,7,8}$ as the right, we could see
that $J_{4,5}$ and $J_{6,7}$ play the role of tunable local $X$ operation
for individual 2-D Hilbert space, and the perpetual Ising type interaction
induce an untunable term of the type $a(\sigma _{L}^{z}+\sigma
_{R}^{z})+b\sigma _{L}^{z}\sigma _{R}^{z}$ in the Hamiltonian.
Unfortunately, techniques developed for NMR such as refocusing usually can
not be employed in other systems (especially many solid state systems
including quantum dot and superconducting circuits) because the assumption
of fast, strong pulse can not be valid, so we choose an alternative way to
perform CPHASE gate.

Starting with the four possible initial states with degenerate static
energy, $\left\{ \left\vert 100\right\rangle ,\left\vert 010\right\rangle
\right\} _{3,4,5}\bigotimes \left\{ \left\vert 010\right\rangle ,\left\vert
001\right\rangle \right\} _{6,7,8}$, in the first step we tune only $J_{4,5}$
while set $J_{6,7}=0$. The $XXZ$ interaction between spin 4 and 5 keeps
states $\left\vert 100001\right\rangle $ and $\left\vert 100010\right\rangle
$ unchanged.

In space $S_{3}$ taking Bloch sphere representation we see the
transformation induced by $J_{4,5}$ is a rotation about $X$ axis. But in
space $S_{4}$ the induced rotation is more complex. We note that since our
initial state in $S_{4}$ is only $\left\vert 010010\right\rangle $, the
induced transformation in $S_{4}$ is finally reduced to a 2-D subspace $%
S_{4}^{^{\prime }}=span\{\left\{ \left\vert 010\right\rangle ,\left\vert
001\right\rangle \right\} _{3,4,5}\bigotimes \left\{ \left\vert
010\right\rangle \right\} _{6,7,8}\}$. Under basis $\left\{ \left\vert
010010\right\rangle ,\left\vert 001010\right\rangle \right\} $ the reduced
Hamiltonian is
\begin{equation}
H_{4}^{^{\prime }}(J_{4,5},0)=\left[
\begin{array}{cc}
0 & 2J_{4,5} \\
2J_{4,5} & -4J_{2}%
\end{array}%
\right]
\end{equation}%
The induced transformation is a rotation about an axis on $X-Z$ plane
because the static energy of $\left\vert 010010\right\rangle \allowbreak $
and $\left\vert 001010\right\rangle $ are slightly different due to the
long-range interaction $H_{L}$.

So in $S_{4}^{^{\prime }}$, with tunable $J_{4,5}$ acting as $\sigma ^{x}$%
-operation, we can perform a transformation $\exp \left\{ -i\pi \sigma
^{x}/2\right\} $, which is exactly a $\pi $ rotation around the $X$-axis in
Bloch sphere representation. We assume that the maximum value of $J_{4,5}$
we could tune to is $X_{1}/2$. We set parameter $\theta $ and $X_{2\text{ }}$%
as $\cos \theta =\frac{X_{1}}{\sqrt{X_{1}^{2}+\left( 2J_{2}\right) ^{2}}}$, $%
\sin \theta =\frac{2J_{2}}{\sqrt{X_{1}^{2}+\left( 2J_{2}\right) ^{2}}}$, $%
\cos 2\theta =\frac{X_{2}}{\sqrt{X_{2}^{2}+\left( 2J_{2}\right)
^{2}}}$, and $\sin 2\theta =\frac{2J_{2}}{\sqrt{X_{2}^{2}+\left(
2J_{2}\right) ^{2}}}$. Then we define rotation (we set $\hbar =1$)
\begin{equation}
R(X)=\exp \left[ iH_{4}^{^{\prime }}(X,0)\pi /2\sqrt{X^{2}+\left(
2J_{2}\right) ^{2}}\right]   \label{rotation1}
\end{equation}%
Thus we have
\begin{equation}
R(X_{1})=i\left[
\begin{array}{cc}
\sin \theta  & \cos \theta  \\
\cos \theta  & -\sin \theta
\end{array}%
\right]   \label{rotation2}
\end{equation}%
\begin{equation}
R(X_{2})=i\left[
\begin{array}{cc}
\sin 2\theta  & \cos 2\theta  \\
\cos 2\theta  & -\sin 2\theta
\end{array}%
\right]   \label{rotation3}
\end{equation}%
And the combined rotation in subspace $S_{4}^{^{\prime }}$ is what we want:
\begin{equation}
R(X_{1})R(X_{2})R(X_{1})=\exp \left\{ -i\pi \sigma ^{x}/2\right\}
\label{rotation4}
\end{equation}

These above tuning of $J_{4,5}$ at last form a unitary transformation $U_{1}$
in the $9$-D space $S$: It is nontrivial only in space $S_{3}$ and $%
S_{4}^{^{\prime }}$: $\left\vert 100001\right\rangle $ and $\left\vert
100010\right\rangle $ remain unchanged; $\left\vert 010010\right\rangle $ is
transformed exactly into $\left\vert 001010\right\rangle $; $\left\vert
010001\right\rangle $ is transformed into superposition of $\left\vert
010001\right\rangle $ and $\left\vert 001001\right\rangle $. As shown in
Eqs. \ref{rotation1}--\ref{rotation4}, the required time for this step is $%
T_{1}=\pi /\sqrt{X_{1}^{2}+\left( 2J_{2}\right) ^{2}}+\pi /2\sqrt{%
X_{2}^{2}+\left( 2J_{2}\right) ^{2}}$. We could see that $T_{1}$
mainly depends the maximal $XY$ interaction strength we could
have. We further note an important fact that if we re-perform the
tuning of $J_{4,5}$ with inverse strength, i.e. $X_{1}$ to
$-X_{1}$, $X_{2}$ to $-X_{2}$, we can get the inverse operation of
$U_{1}$ in $S$.

In the second step we tune $J_{4,5}$ to zero but come to control $J_{6,7}$.
Quite similar to the previous step, this exchange interaction keeps states
in $S_{1}$ and $S_{3}$ unchanged while induce transformations in $S_{2}$ and
$S_{4}$. In $S_{2}$ under Bloch sphere representation the induced
transformation is a rotation about $X$ axis, while in subspace $S_{4}$,
since the initial state $\left\vert 010010\right\rangle $ is exactly
transformed into state $\left\vert 001010\right\rangle $ by the first step,
the induced transformation by $J_{6,7}$ in $S_{4}$ is restricted to a 2-D
subspace $S_{4}^{^{\prime \prime }}=\left\{ \left\vert 001\right\rangle
\right\} _{3,4,5}\bigotimes \left\{ \left\vert 010\right\rangle ,\left\vert
100\right\rangle \right\} _{6,7,8}$. Under basis $\left\{ \left\vert
001010\right\rangle ,\left\vert 001100\right\rangle \right\} $ the reduced
Hamiltonian is
\begin{equation}
H_{4}^{^{\prime \prime }}(0,J_{6,7})=\left[
\begin{array}{cc}
-4J_{2} & 2J_{6,7} \\
2J_{6,7} & -4J_{1}%
\end{array}%
\right]
\end{equation}%
The form of $H_{4}^{^{\prime \prime }}(0,J_{6,7})$ is quite
similar to that of the previous $H_{4}^{^{\prime }}(J_{4,5},0)$,
so we can perform another unitary transformation $U_{2}$ just
similar to the first step which implement a $\pi $ rotation around
the $X$-axis in space $S_{4}^{^{"}}$, transforming state
$\left\vert 001010\right\rangle $ to $\left\vert
001100\right\rangle $.

After the above two steps, we review the intermediate states we get: $%
\left\vert 100001\right\rangle $ remain unchanged; $\left\vert
100010\right\rangle $ is changed into the superposition of
$\left\vert 100010\right\rangle $ and $\left\vert
100100\right\rangle $; $\left\vert 010001\right\rangle $ is
changed into the superposition of $\left\vert 010001\right\rangle
$ and $\left\vert 001001\right\rangle $; $\left\vert
010010\right\rangle $ is changed into $\left\vert
001100\right\rangle $. The
previous three intermediate states are degenerate under Hamiltonian $%
H_{M}\left( 0,0\right) $ but the last state has nonzero static energy $%
-4J_{1}$. Therefore, in the third step we tune off all exchange
coupling for a time interval $\tau $. In this period the state
$\left\vert 001100\right\rangle $ experience an additional phase
$\varphi $ due to its non-zero static energy. In the last step we
can perform the inverse operation of $U_{2}$ and $U_{1}$ to
transform the four intermediate states back to the initial four
states.

After all the above operations we have performed a CPHASE gate $U\left(
\varphi \right) $ between two qubits, adding a controllable phase $\varphi $
on the state $\left\vert 010010\right\rangle $ while remaining other states
unchanged. As we mentioned before, we use spin states $\left\vert
01\right\rangle $ for spin 3-4 and spin 7-8 as logical $\left\vert
0\right\rangle $ and $\left\vert 10\right\rangle $ as logical $\left\vert
1\right\rangle $, in this 4-D representation the gate $U\left( \varphi
\right) $\ we obtain is
\begin{equation}
\left[
\begin{array}{cccc}
1 & 0 & 0 & 0 \\
0 & e^{i\varphi } & 0 & 0 \\
0 & 0 & 1 & 0 \\
0 & 0 & 0 & 1%
\end{array}%
\right]
\end{equation}

After implementation of the CPHASE gate $U$, we note that \cite%
{Zhou3}  $\left( \sigma _{2}^{x}\ast U(2\varphi )\right)
^{2}=e^{i\varphi }e^{i\sigma _{1}^{z}\varphi }$, thus with the single qubit $%
\sigma ^{x}$ rotation of the second qubit and CPHASE gate between the first
and the second qubit, the single qubit $\sigma ^{z}$ rotation of the first
qubit is obtained.

In summary of this section, we have demonstrated all the required
universal gates for QC. In this scheme, by using appropriate
encoding method, influence of the next-nearest-neighbor
interaction is effectively ruled out, thus the precision of
quantum gates updated. Besides, the speed of the CPHASE gate
mainly depends on the strength of nearest-neighbor exchange
interaction\textit{, }it is not restricted by the small value of
$J_{2}$. The quantum information speed of this scheme is in the
same level with that of previous schemes in which the
next-nearest-neighbor interaction was neglected.

\section{A potential physical realization: Josephson Junction Charge Qubit}

Now we consider  the application of our scheme to realistic
systems. We consider the long-range interaction in Josephson
charge qubit system as an
example. The typical Josephson-junction charge qubit is shown in Fig. 3 \cite%
{Schon1}: It consists of a small superconducting island with $n$ excess
Cooper pairs, connected by a tunnel junction with capacitance $C_{J}$ and
Josephson coupling energy $E_{J}$ to a superconducting electrode. A control
gate voltage $V_{g}$ is coupled to the system via the gate capacitor $C_{g}$%
. The Hamiltonian of the Cooper pairs box (CPB) is
\begin{equation}
H_{Josephson}=E_{c}(n-n_{g})^{2}+E_{J}\cos \Phi
\end{equation}%
where $E_{c}=(2e)^{2}/2\left( C_{g}+C_{J}\right) $ is the charge energy, $%
n_{g}=C_{g}V_{g}/2$ is the gate charge, and $\Phi $ is the conjugate
variable to $n$. When $E_{c}\gg E_{J}$, by choosing $n_{g}$ close to the
degeneracy point $n_{g}=1/2$, only the states with $0$ and $1$ Cooper pairs
play a role while all other charge states in much higher energy level can be
ignored. In this case the CPB can be reduced to an effective two-state
quantum system. A further step is replacing the single Josephson junction by
two identical junctions in a loop configuration in order to gain tunable
tunnelling amplitude \cite{Schon2}. By making replacement $n=\left( 1+\sigma
^{z}\right) /2$, the effective Hamiltonian can be written in the spin-$1/2$
notation as
\begin{equation}
H_{Josephson}=-B_{z}\sigma ^{z}-B_{x}\sigma ^{x}
\end{equation}%
where the state with $0$ Cooper pairs corresponds to the spin state $%
\left\vert \downarrow \right\rangle $ and $1$ Cooper pairs to $\left\vert
\uparrow \right\rangle $. $B_{z}$ and $B_{x}$ are the effective magnetic
fields which are controlled by the biased gate voltage and frustrated
magnetic flux.

\begin{figure}[tbh]
\epsfig{file=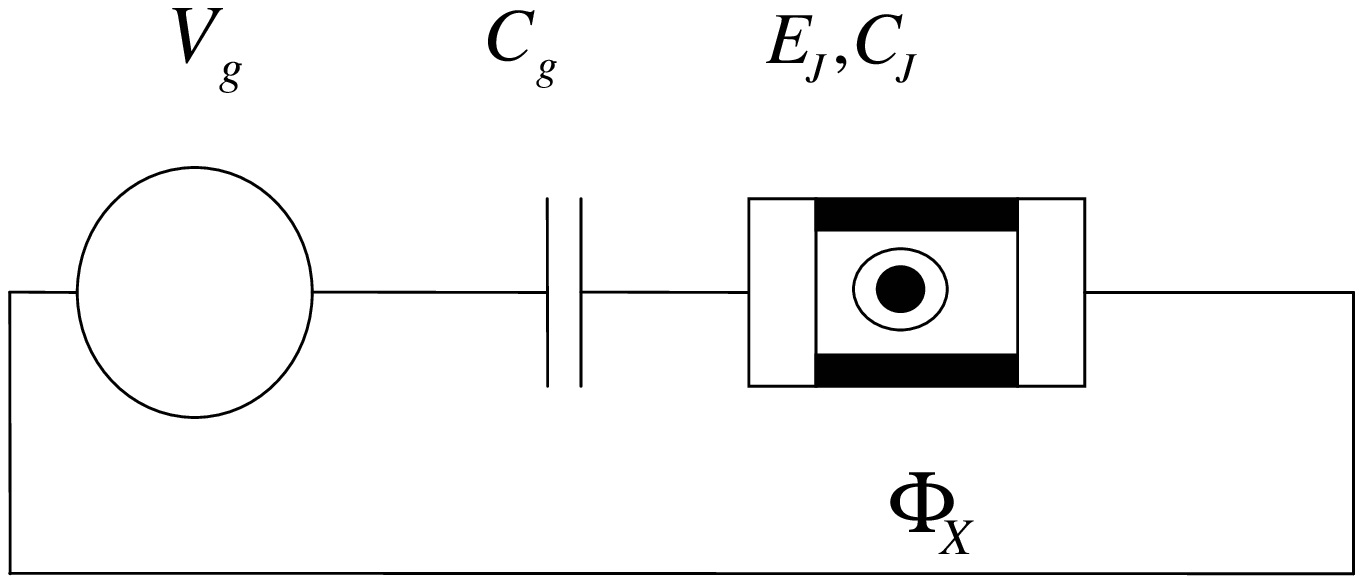,width=8cm}
\caption{A Josephson charge qubit formed by a superconducting single-charge
box and a tunnel junction.The single Josephson junction is replaced by a
flux-frustrated SQUID so that the effective Josephson coupling is tunable.}
\end{figure}

For coupling two CPBs, the direct capacitance coupling \cite{Nakamura2}\cite%
{Nakamura3} is most intrinsic, but its drawback is also obvious, that is,
the coupling strength induced by connective capacitances is untunable. For a
system consist of $M$ CPBs coupled with each other by capacitances, the
static charge energy can be written as \cite{Mooij2}\cite{Berman}%
\begin{equation}
H_{C}=\frac{\left( 2e\right) ^{2}}{2}\overrightarrow{n}C^{-1}\overrightarrow{%
n}^{\dagger }  \label{CapacitanceHamiltonian}
\end{equation}%
where $\overrightarrow{n}%
=(n_{1}-n_{g1},n_{2}-n_{g2},n_{3}-n_{g3},......n_{M}-n_{gM})$ is the charge
number vector of the $M$ CPBs, and $C$ is the capacitance matrix of the
system whose diagonal term $C_{i,i}$ equals to the sum of capacitance around
the $ith$ CPB, and non diagonal term $C_{i,j}$ corresponds to the connective
capacitance between CPB $i$ and $j$.

A schematic plot of an array of $N$ capacitively coupled CPBs is shown in
Fig. 4. The CPBs have Josephson energies $E_{Ji}$ and capacitances $C_{Ji}$.
Each CPB is connected to the control gate voltages $V_{gi}$ via a gate
capacitance $C_{gi}$. The $i$th intermediate CPB is connected to its
neighboring $\left( i\pm 1\right) $th CPBs via the connective capacitors $%
C_{ci,i\pm 1}$. We assume that all CPBs are identical with the same
capacitances: $C_{gi}=C_{g}$, $C_{Ji}=C_{J}$, and all the coupling
capacitance $C_{ci,i+1}$ are equal to $C_{c}$. So we have a tridiagonal
capacitance matrix for this system. For the intermediate qubits of the
array,
\begin{equation}
C_{i,j}=C_{0}[\delta _{i,j}(1+2\epsilon )-\delta _{i,j\pm 1}\epsilon
],1<i,j<N  \label{CapacitanceMatrix}
\end{equation}%
where $C_{0}=C_{g}+C_{J}$ and $\epsilon =C_{c}/C_{0}$. For qubits on the
edge of the array a small correction is needed: $C_{1,1}=C_{N,N}=C_{0}(1+%
\epsilon )$ while $C_{1,2}=C_{N-1,N}=-C_{0}\epsilon $.

\begin{figure}[tbh]
\epsfig{file=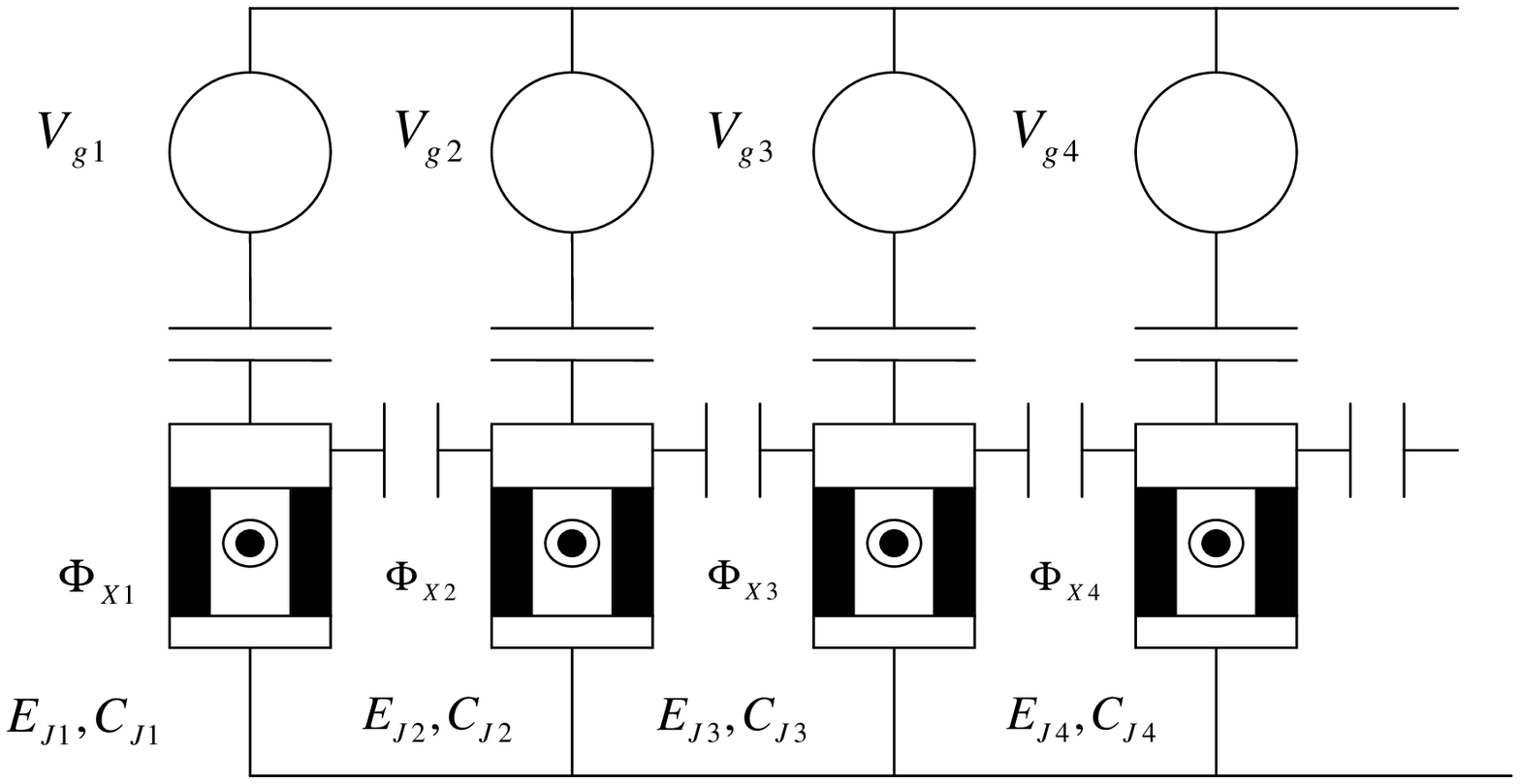,width=8cm}
\caption{A schematic illustration of array of capacitively coupled CPBs.}
\end{figure}

Since $C$ is a tridiagonal matrix, $C^{-1}$ has nonzero matrix
elements on the second, third and other diagonals which
characterize the capacitance induced Coulomb interaction between
different CPBs. If $\epsilon \ll 1$, the off-diagonal elements of
$C^{-1}$ decay exponentially as $C_{i,j}^{-1}\sim
C_{i,i}^{-1}\epsilon ^{\left\vert i-j\right\vert }$. So the
influence of the
long-range interaction can be reduced by taking the coupling capacitances $%
C_{c}$ much smaller than the on-site capacitances $C_{0}$. Again by making
replacement $n_{i}=\left( 1+\sigma _{i}^{z}\right) /2$, we get that the
interaction term $C_{i,j}^{-1}(n_{i}-n_{gi})(n_{j}-n_{gj})$ provides
always-on Ising type interaction between CPB $i$ and $j$.

If we go further, replacing the coupling capacitance by SQUID, we can have
tunable $XY$ type exchange interaction between neighboring qubit besides
perpetual Ising type interaction (Fig. 5) \cite{Siewert1}\cite{Siewert2}\cite%
{Sun1}. Due to flux quantization the phase across the coupling SQUID is $%
\Phi _{1}-\Phi _{2}+\alpha $, $\alpha $ being a constant controlled by the
frustrated flux. If we tune $\alpha $ to zero, the coupling term $\cos (\Phi
_{1}-\Phi _{2})$ thus induce $XY$ type exchange interaction $\sigma
_{1}^{x}\sigma _{2}^{x}+\sigma _{1}^{y}\sigma _{2}^{y}$.

\begin{figure}[tbh]
\epsfig{file=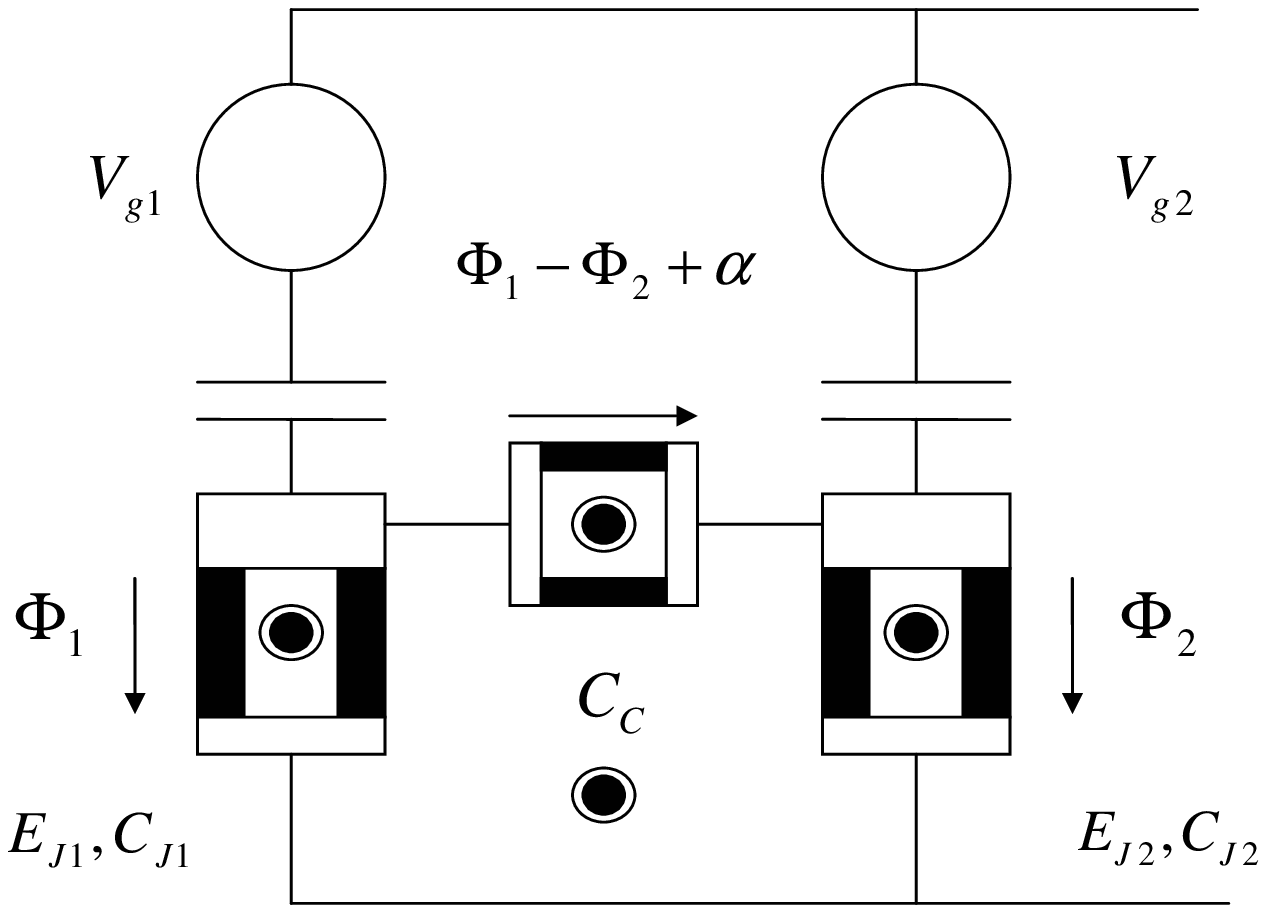,width=8cm}
\caption{A schematic plot of two CPBs coupled by a SQUID.}
\end{figure}

Thus we can see the correspondence between the theoretical
Hamiltonian in Eq. \ref{modelhamiltonian1} and the realistic
physical system: the tunable SQUIDs of single qubits induce
tunable $\sigma ^{x}$ terms in the single qubit part of
Hamiltonian; all the bias voltages are biased on the degeneracy
point $V_{gi}=1/2$ in order to prevent the qubits from the $1/f$
noise effect \cite{Vion}, implying that there is no $\sigma ^{z}$
term in the single qubit part of Hamiltonian; the exponentially
decay capacitance coupling corresponds to the nearest Ising
interaction part in $H_{I}$ and the next-nearest-neighbor Ising
interaction part $H_{L}$, while the tunable SQUID coupling
corresponds to the tunable $XY$ interaction part in $H_{I}$. We
also see that the required performances in the above scheme just
correspond to tuning external magnetic field frustrated in the
SQUID loops.

By taking $\epsilon \ll 1$ the effect of the long-range
interaction in Josephson charge qubit array system can be reduced
but can never be negated. Moreover, the speed of two qubits
operation depends on $\epsilon $, decreasing the value of
$\epsilon $ would slow the speed of two qubits gate. Besides,
making $\epsilon $ smaller and smaller may be highly challenging
in experimental realization. So we may choose an alternative way
to handle the problem of always-on non-nearest-neighbor coupling.
The main idea is that, based on the decay property of the
long-range interaction, we can take some lower order terms of
long-range interaction into account while taking the higher order
terms as random noise. Thus we may use the encoding schemes
discussed in Sec. III to negate the influence of the few lowest
order couplings.

Schemes of using capacitively coupled Josephson array to perform QC have
been proposed \cite{Berman}\cite{Averin}. Our scheme offer an alternative
idea to handle the long-range interaction problem. The only parameter
required to tune is the flux frustrated in SQUID loops, while the gate
voltages are frozen on the degeneracy point, which prevents qubits from
severe dephasing.

It should be noted that, in our scheme although the
next-nearest-neighbor interaction effect is entirely negated, the
higher order long-range couplings still work. A natural
generalization is using $3$ spins as one \textquotedblleft
blockade\textquotedblright\ to negate the influence of the third
order long-range interaction (first order being the
nearest-neighbor coupling while second order being the
next-nearest-neighbor coupling), or even $m$ spins as one
\textquotedblleft blockade\textquotedblright\ to negate the
influence of the $m$th order long-range interaction. Following
analysis similar to the estimation in appendix. A, we could
conclude that, if we have negated the influence of the $m$th order
interaction, the speed of deviation from ideal unitary evolution
will be $O(nJ_{m+1})$, $J_{m+1}$ being the characteristic
interaction strength of the $(m+1)$th order coupling. In $1$-D
Josephson charge qubit array system the strength of the long-range
interaction have exponentially decay property, this means that,
for a certain $n$, we could always make $nJ_{m+1}$ small enough.
So we could apply the generalization of our QC scheme to this
physical system to suppress the speed of deviation induced by
long-range interaction into some tolerable domain. Another
potential extension is generalizing the nearest-neighbor coupling
in our scheme from $XXZ$ type to various other types including
$XY$ type and Heisenberg type. We could also translate our idea of
handing long-range interaction in this paper to other
implementations.

\section{conclusion}

In conclusion, in this paper we have studied the non-nearest-neighbor
interaction effect in $1$-D spin-$1/2$ chain model. We prove that the
quantum gate deviation induced by the long-range interaction may challenge
the scalability of quantum computing. We further propose a QC scheme in
order to suppress influence of long-range interaction. In this scheme, using
appropriate encoding method, we effectively neutralize influence of the
next-nearest-neighbor interaction, thus the precision of quantum gates
updated. The quantum information speed of this scheme is in the same level
with that of previous schemes in which the long-range interaction strength
was ignored. We also discuss the feasibility of the scheme in $1$-D
Josephson charge qubit array system. This scheme may offer improvement in
dealing with systematic errors in scalable quantum computing.

\begin{acknowledgments}
Y. Hu thanks J. M. Cai, M. Y. Ye, X, F, Zhou, and Y. J. Han for fruitful
discussion. This work was funded by National Fundamental Research Program
(2001CB309300), the Innovation funds from Chinese Academy of Sciences,
NCET-04-0587, and National Natural Science Foundation of China (Grant No.
60121503, 10574126).
\end{acknowledgments}

\appendix

\section{Estimation of deviation induced by long-range interaction}

\label{appendix:resonator}

Let us follow schemes developed in ref. \cite{Bose1}. As shown in Fig. 1: In
the spin-$1/2$ chain, only the even spins (the hexagonal ones) are chosen as
qubits while the odd spins (the rounded ones) are used as \textquotedblleft
blockade\textquotedblright .

When we perform quantum gates following schemes developed in ref. \cite%
{Bose1}, the theoretical expected unitary evolution is $U=\exp
(-itH_{Ideal}) $ while the realistic evolution is $V=\exp
[-it(H_{Ideal}+H_{R})]$, We could estimate the deviation $\left\Vert
U-V\right\Vert $ in single and multi qubit operations:

\emph{(1) Idle. }When the whole chain is in idle status, the single qubit
part $H_{S}$ and the exchange interaction term in $H_{I}$ are all tuned off,
the $n+1$ odd spins used as \textquotedblleft blockade
spins\textquotedblright\ are \textquotedblleft frozen\textquotedblright\ in
definite states $\left\vert 0\right\rangle $ or $\left\vert 1\right\rangle $%
. Since the permanent nearest-neighbor and next-nearest-neighbor Ising type
interaction do not transfer energy, we could reduce our discussion in a $%
2^{n}$-D Hilbert space $H$ which is direct product of all the $n$ qubits.
The expected evolution is $U_{Idle}=I$. But the realistic evolution is
\begin{equation}
V_{Idle}=\exp \{-iJ_{2}t[-n+\sum_{i=1}^{n-1}\sigma _{2i}^{z}\sigma
_{2i+2}^{z}]\}  \label{RealIdle}
\end{equation}%
$\sigma _{j}^{x}$, $\sigma _{j}^{y}$, and $\sigma _{j}^{z}$ being Pauli
operators of the $j$th spin. Eigenvalues of $V_{Idle}$ varys from $\exp
\{iJ_{2}t\}$ to $\exp \{-iJ_{2}t[-2n+1]\}$. $\exp \{iJ_{2}t\}$ corresponds
to states that the $n$ qubits are all in state $\left\vert 0\right\rangle $
or all in state $\left\vert 1\right\rangle $, while $\exp \{-iJ_{2}t[-2n+1]\}
$ corresponds to states that for any $i$ the $2i$th spin and the $2(i+1)$th
spin are in opposite states. We could choose a proper energy zero point for
realistic evolution process, this means we could add a proper phase factor
on $V_{Idle}:$ $V_{Idle}\rightarrow e^{i\varphi }V_{Idle}.$ But no matter
how we choose the energy zero point, we could prove that, for any $\varphi $%
, either
\begin{equation}
\left\vert e^{i\varphi }\exp \{-iJ_{2}t[-2n+1]\}-1\right\vert \geq
\left\vert \exp \{-iJ_{2}t[n-1]\}-1\right\vert   \label{Idle eigenvalue1}
\end{equation}%
or
\begin{equation}
\left\vert e^{i\varphi }\exp \{iJ_{2}t\}-1\right\vert \geq \left\vert \exp
\{-iJ_{2}t[n-1]\}-1\right\vert   \label{idle eigenvalue2}
\end{equation}%
is valid. So we could give an estimation of deviation for idle status:
\begin{equation}
\left\Vert U-V\right\Vert _{Idle}\geq \left\vert \exp
\{-iJ_{2}t[n-1]\}-1\right\vert   \label{IdleProve}
\end{equation}

\emph{(2) Performing }$\sigma ^{z}$\emph{\ rotations. }When performing $%
\sigma ^{z}$ rotations on the $2i$th spin, parts for other spins in $H_{S}$
and all $J_{i,i+1}$ for any $i$ are tuned off, the $n+1$ odd spins used as
\textquotedblleft blockade spins\textquotedblright\ are \textquotedblleft
frozen\textquotedblright\ in definite states $\left\vert 0\right\rangle $ or
$\left\vert 1\right\rangle $ as shown in Fig. 1. We could estimate the
deviation $\left\Vert U_{z}-V_{z}\right\Vert $ in a $2^{n-1}$-D Hilbert
space $H_{z}$. $H_{z}$ is defined as below: $H_{z}$ is a subspace of $H$. It
is the direct product of all the $n$ qubits except the $i$th, and for any
state in $H_{z}$, the $i$th qubit is in state $\left\vert 0\right\rangle $.
The expected evolution in $H_{z}$ is $U_{z}=\exp [-itB_{z}^{2i}]$. But the
realistic evolution is
\begin{equation}
V_{z}=\exp \{-it[B_{z}^{2i}-nJ_{2}+J_{2}\sigma _{2i-2}^{z}+J_{2}\sigma
_{2i+2}^{z}+J_{2}\sum_{k=1,k\neq i-1,k\neq i}^{n-1}\sigma _{2k}^{z}\sigma
_{2k+2}^{z}]\}  \label{Zreal}
\end{equation}%
Thus the estimation of deviation in $H_{z}$ is quite similar to previous
idle status:
\[
\left\Vert U-V\right\Vert _{z}^{H_{z}}\geq \left\vert \exp
\{-iJ_{2}t[n-1]\}-1\right\vert
\]%
We note that the norm of operator $U-V$ in the whole space $H$ could not be
smaller than the norm of operator $U-V$ reduced in a subspace $H_{z}$ of $H$%
. So we give an estimation of deviation for performing $\sigma ^{z}$\emph{\ }%
gate:
\begin{equation}
\left\Vert U-V\right\Vert _{z}\geq \left\vert \exp
\{-iJ_{2}t[n-1]\}-1\right\vert
\end{equation}

\emph{(3) Performing }$\sigma ^{x}$\emph{\ rotations. }Discussion of\emph{\ }%
performing $\sigma ^{x}$ rotations on the $i$th qubit is similar to previous
discussion of performing $\sigma ^{z}$ rotations on the $i$th qubit.
Similarly we could define a subspace $H_{x}$ of $H$: It is the direct
product of qubits except the $(i-1)$th, $i$th, and $(i+1)$th qubits, and for
any state in $H_{x}$, the $(i-1)$th qubit is in state $\left\vert
0\right\rangle ,$ the ith qubit is in state $(\left\vert 0\right\rangle
+\left\vert 1\right\rangle )/\sqrt{2},$ and the $(i+1)$th qubit is in state $%
\left\vert 1\right\rangle .$ So we have $J_{2}\sigma _{2i-2}^{z}\sigma
_{2i}^{z}+J_{2}\sigma _{2i}^{z}\sigma _{2i+2}^{z}=0.$ The expected evolution
in this subspace is $U_{x}=\exp [-itB_{x}^{2i}]$, while the realistic
evolution is
\begin{equation}
V_{x}=\exp \{-it[B_{x}^{2i}-nJ_{2}+J_{2}\sigma _{2(i-2)}^{z}-J_{2}\sigma
_{2(i+2)}^{z}+J_{2}(\sum_{k=1}^{i-2}+\sum_{k=i+1}^{n-1})\sigma
_{2i}^{z}\sigma _{2i+2}^{z}]\}  \label{XReal}
\end{equation}%
Eq. \ref{XReal} is similar to Eq. \ref{Zreal}. We give the estimation:
\begin{equation}
\left\Vert U-V\right\Vert _{x}\geq \left\vert \exp
\{-iJ_{2}t[n-3]\}-1\right\vert
\end{equation}

\emph{(4) Performing inter-qubit gate. }Inter-qubit gate between the $i$th
qubit and the $(i+1)$th qubit is achieved by tuning the inter-spin exchange
interaction strength $J_{2i,2i+1}$ and $J_{2i+1,2i+2}$ while other exchange
interaction terms are all tuned off. Suppose in idle status the blockade
spin $2i$ is in state $\left\vert 0\right\rangle $, we could estimate the
deviation in a $2^{n-2}$ Hilbert space $H_{int}.$ $H_{int}$ is defined as
follow: $H_{int}$ is a subspace of $H$; It is the direct product of all
qubits except the $i$th and $(i+1)$th qubits. For any state in $H_{int}$,
the $i$th and the $(i+1)$th qubits are both in state $\left\vert
0\right\rangle $. Obviously the ideal expected evolution in this space is $I$%
, but the realistic evolution is%
\begin{equation}
V_{int}=\exp \{-it[-(n-1)J_{2}-J_{2}\sigma _{2(i-1)}^{z}-J_{2}\sigma
_{2(i+2)}^{z}+J_{2}(\sum_{k=1}^{i-2}+\sum_{k=i+2}^{n-1})\sigma
_{2k}^{z}\sigma _{2k+2}^{z}]\}
\end{equation}%
Quite similar to previous estimations we give:\emph{\ }%
\begin{equation}
\left\Vert U-V\right\Vert _{int}\geq \left\vert \exp
\{-iJ_{2}t[n-2]\}-1\right\vert
\end{equation}


\end{document}